# The structure of $PX_3$ (X=Cl, Br, I) molecular liquids from X-ray diffraction, Molecular Dynamics simulations and Reverse Monte Carlo modeling

Szilvia Pothoczki,[a)] László Temleitner, and László Pusztai

*Institute for Solid State Physics and Optics, Wigner Research Centre for Physics,*

*Hungarian Academy of Sciences, Konkoly-Thege M. út 29-33, 1121, Budapest, Hungary*

**Abstract**

Synchrotron X-ray diffraction measurements have been conducted on liquid phosphorus trichloride, tribromide and triiodide. Molecular Dynamics simulations for these molecular liquids were performed with a dual purpose: (1) to establish whether existing intermolecular potential functions can provide a picture that is consistent with diffraction data; (2) to generate reliable starting configurations for subsequent Reverse Monte Carlo modelling. Structural models (i.e., sets of coordinates of thousands of atoms) that were fully consistent with experimental diffraction information, within errors, have been prepared by means of the Reverse Monte Carlo method. Comparison with reference systems, generated by hard sphere-like Monte Carlo simulations, was also carried out to demonstrate the extent to which simple space filling effects determine the structure of the liquids (and thus, also estimating the information content of measured data). Total scattering structure factors, partial radial distribution functions and orientational correlations as a function of distances between the molecular centres have been calculated from the models. In general, more or less antiparallel arrangements of the primary molecular axes that are found to be the most favourable orientation of two neighbouring molecules. In liquid $PBr_3$ electrostatic interactions seem to play a more important role in determining intermolecular correlations than in the other two liquids; molecular arrangements in both $PCl_3$ and $PI_3$ are largely driven by steric effects.

[a)]Author to whom correspondence should be addressed. Electronic mail: pothoczki.szilvia@wigner.mta.hu.

# I. INTRODUCTION

One of the most interesting issues in connection with the structure of the molecular liquids is how, and to what extent, the molecular shape determines the mutual orientations of molecules, i.e. the orientational correlations. Furthermore, if the molecules possess permanent dipole moments, how do dipolar interactions influence the short-range ordering. The competition (and therefore, eventually, a delicate balance) between electrostatic (dipolar) and steric effects, which can be different from system to system, prodded us to investigate a peculiar group of molecular liquids, namely the $PX_3$ (X=Cl, Br, I) family. Another reason why these materials deserve attention is that they are analogues to well-known tetrahedral liquids such as carbon tetrachloride, since the lone electron pair of the P atom forces the molecular shape to be a distorted tetrahedron[1].

Structural studies only for liquid $PBr_3$ have been published[2,3] previously. Results from neutron diffraction experiments[2] and from Reverse Monte Carlo (RMC) modeling[3] (based on neutron diffraction data only) motivate a more extensive and systematic investigation. The latter study, for instance, revealed that some of the simulation parameters, such as the minimum distance allowed between two atoms of neighbouring molecules ('intermolecular cut-off distance' in the RMC terminology), may influence (or indeed, determine) angular correlations.

A significant step towards clarifying the pending issues would be to perform potential based Molecular Dynamics (MD) simulations[4] of all liquids in question, for several reasons: (1) comparing the structure factor obtained from the MD trajectory with the experimental one provides valuable information concerning the trustworthiness of the intermolecular potential functions; (2) the final configuration of an MD simulation is a suitable initial configuration for a subsequent Reverse Monte Carlo[5] calculation[6]. Here we intend to follow a uniform 'protocol' for each $PX_3$ liquid considered here: (1) determine (or, if possible, find in the literature) the total scattering structure factor, $F(Q)$[7]; (2) perform MD simulation (to the best of our knowledge, this is the first attempt to perform potential based computer simulations for these materials); (3) apply the Reverse Monte Carlo technique of structural modelling so that particle configurations consistent with measured diffraction data may be produced. In order to monitor the influence of purely steric effects we also utilize a 'hard sphere model' (HS), defined only by the atomic packing and the molecular shape. This approach proved to be quite instructive in our earlier studies.[8,9]

In order to be able to provide new 'structural models' (i.e., coordinates of thousands of particles) via Reverse Monte Carlo (RMC), new X-ray diffraction data have been taken for $PCl_3$, $PBr_3$ and $PI_3$ liquids. Further, we felt it timely to reconsider the RMC simulation of $PBr_3$ molecular liquid[2] in view of the present MD simulations and the new X-ray data set.

The total scattering structure factors, partial radial distribution functions (prdf) and two different distance-dependent orientational correlations functions (see, e.g., Ref. 10) are calculated from each structural model generated here ($PBr_3$_MD, $PBr_3$_RMC, $PBr_3$_HS, $PCl_3$_MD, $PCl_3$_RMC, $PCl_3$_HS, $PI_3$_MD, $PI_3$_RMC, $PI_3$_HS; for a detailed specification of these structural models, see below). For a proper understanding of the main issue of this work, the interplay between steric and electrostatic effects, an atomic level structural comparison between the three liquids ($PBr_3$, $PCl_3$, $PI_3$) is provided, taking into account the properties of the different type of simulation methods.

The paper is organized as follows: Section 2 describes the new X-ray diffraction experiments. In Section 3 details of the MD simulation and the RMC modeling, as well as the definition of the orientational correlation functions applied here can be found. Section 4 describes the main findings and finally, in Section 5, concluding remarks are given.

## II. EXPERIMENTAL DETAILS

$PI_3$ and $PBr_3$ were purchased from Aldrich while $PCl_3$ was from Wako Chemicals; the purity was 99% for all materials. The samples were sealed in 2 mm diameter thin-walled quartz capillaries (made by Glass Müller) as follows. Under static argon atmosphere in a glovebox, the liquid samples ($PCl_3$ and $PBr_3$) were transferred to the capillaries by glass Pasteur pipettes. $PI_3$, that is solid at room temperature, was milled in an agate mortar before transfer. The capillaries were temporarily sealed with vacuum grease, then removed from the glovebox and closed by an $O_2$ torch under streaming argon gas coming from an upside-down funnel. The procedure was completed within 5 minutes after removing the samples from the glovebox.

X-ray diffraction data were collected using the single-detector diffractometer setup of the BL04B2 (high-energy X-ray diffraction) beamline[11] located at the SPring-8 synchrotron radiation facility (Hyogo, Japan). This instrument is optimized for the study of amorphous and liquid samples. The intensity of the incident X-rays was monitored by an ionization chamber filled with Ar gas and the scattered X-rays were detected by a HPGe detector. The incident X-ray wavelength of 0.2021 Å (corresponding to the energy of 61.34 keV) was selected as a compromise between the available Q–range, beam intensity and absorption of the materials. This setup provided an accessible Q range of 0.3–24 Å$^{-1}$.

The liquid $PCl_3$ and $PBr_3$ samples were measured under ambient conditions, using the automatic sample changer. The $PI_3$ sample was placed into a furnace available at the beamline and measured at 80.5(±1.5)°C (353.7 K), well above its melting point (61°C)[12]. Before the start of the measurement, gas-bubbles were removed by shaking the capillary at the target temperature.

The data were corrected for scattering from background and the empty capillary, sample self-attenuation and Compton scattering, using standard procedures.[13] As a result, total scattering structure factors[7] have been obtained. For the iodine atoms, where the absorption is non-negligible, the anomalous x-ray form-factor[14] was also taken into account during data evaluation.

For $PCl_3$ and $PBr_3$ the total scattering structure factor is nearly flat beyond 16 $Å^{-1}$, thus this region was excluded from the later RMC modelling. In the case of $PI_3$, due to the very strong absorption of iodine and the presence of the heating device during the measurement, the meaningful total scattering structure factor reached up to only 14 $Å^{-1}$. As it could be shown earlier[15], in quite a few cases even shorter structure factors have proven to be sufficient for a full recovery of the microscopic structure, provided that suitable inverse methods are applied during data evaluation and interpretation; as it is evidenced below, liquid phosphorus triiodide also belongs to this class of materials.

Details of the neutron diffraction experiment for $PBr_3$ can be found in Ref. 2.

## III. COMPUTATIONAL DETAILS

### A. Molecular Dynamics Simulations

Molecular dynamics simulations have been performed in the NVT ensemble with the GROMACS simulation package (version 4.0)[16] at T = 293 K for $PBr_3$ and $PCl_3$, as well as at T = 354 K for $PI_3$ to correspond with experiments. The OPLS all-atom force field[17] was used for all liquids. The LJ parameters and partial charges are given in Table I. The calculation of the non-bonded interactions was optimized by a grid-based neighbor list algorithm updated every 10 steps. Cutoffs for the Coulomb and for the Van der Waals interactions were set at 9 and 10Å; these choices are arguably adequate in view of the small charges.

TABLE I. Lennard-Jones parameters and partial charges for the atom types used in the MD simulations.

|    | σ(Å) | ε(kJ/mol) | Q(e) |
|----|------|-----------|------|
| P  | 3.74 | 0.8368    | 0.213 ($PCl_3$), 0.126($PBr_3$), 0.09($PI_3$) |
| Cl | 3.4  | 1.2552    | -0.071 |
| Br | 3.47 | 1.96648   | -0.042 |
| I  | 3.67 | 2.42672   | -0.03 |

In each calculation, 2000 molecules were put in a cubic simulation cell. Covalent bonds were constrained to their equilibrium values by using the LINCS[18] algorithm, enabling a 2 fs time step. Bond length parameters, as well as atomic number densities (which correspond to the experimental values) are listed in Table II.

The leapfrog algorithm was used for integrating Newton's equations of motion. Temperature was controlled by a Berendsen[19] thermostat with the temperature coupling time constant $\tau$ set to 0.1 ps. The total simulation time was 2000 ps (=2ns) for each liquid.

TABLE II. Initial bond length parameters and atomic number densities used in the MD calculations.

|  | $PCl_3$ | $PBr_3$ | $PI_3$ |
|---|---|---|---|
| P-X (Å) | 2.04 | 2.22 | 2.55 |
| X-X (Å) | 3.12 | 3.43 | 3.85 |
| Atomic number density (Å$^{-3}$) | 0.02751 | 0.02556 | 0.02146 |
| Box length (Å) | 66.2546 | 67.8958 | 71.9727 |

A steepest-descent gradient method was applied prior to the simulations for energy minimization and to avoid atomic overlaps in the systems. In all cases, the total energy reached its equilibrium value within 100 ps.

Partial radial distribution functions, and X-ray and neutron (only for $PBr_3$) weighted total structure factors were calculated using the last 1500 ps of the trajectory. For this purpose the g_rdf software of the GROMACS[16] package was modified[6] concerning the histogram calculation and the Fourier transformation, according to the Reverse Monte Carlo code.[20] Atomic coordinates were stored every 20 ps for the purpose of analyzing orientational correlations (see Section III.c). Results are based on averages over 50 independent configurations for each system.

**B. Reverse Monte Carlo Modeling ('Refinement')**

Reverse Monte Carlo modeling, which may be considered as a 'refinement' to MD results (see, e.g., Ref. 6), was started from initial configurations derived from Molecular Dynamics simulations (see III.A). The RMC method is a way to generate structural models that are fully consistent with results of diffraction experiments within their uncertainties. The principles of RMC modeling have been described elsewhere (e.g. Refs. [5,20-23]), so in this section we concentrate on issues relevant for the systems studied here.

Table III. contains the *fnc* ('fixed neighbors constraints', see Ref. 22) limits and *cut-off* values

used in the RMC runs. The *fnc* are simple neighbor lists that operate via tolerance distances for each intramolecular atomic pair in order to keep the molecules together. The cut-off values in the RMC terminology are minimum atom-atom distances that prevent overlap between molecules (see above). Other properties, such as atomic number densities and simulation box lengths, can be found in Table II since the system sizes in MD and RMC were identical. For liquid $PBr_3$ X-ray and neutron diffraction data sets have been taken into account simultaneously whereas for the other two liquids ($PCl_3$ and $PI_3$) RMC calculations were driven by the new X-ray diffraction data.

TABLE III. Some characteristics of the liquids studied and their computer models. X: Cl ($PCl_3$), Br ($PBr_3$), I ($PI_3$).

|  | $PCl_3$ | $PBr_3$ | $PI_3$ |
|---|---|---|---|
| fnc(P-X) (Å) | 1.99-2.09 | 2.17-2.27 | 2.5-2.6 |
| fnc(X-X) (Å) | 3.075-3.175 | 3.376-3.476 | 3.8-3.9 |
| cut-off P-P (Å) | 4.0 | 4.0 | 4.2 |
| cut-off P-X (Å) | 3.0 | 3.2 | 3.2 |
| cut-off X-X (Å) | 3.2 | 3.3 | 3.3 |

In order to highlight certain structural features, including orientational correlations that arise even with random packing of molecules, every RMC refinement calculation was accompanied by a corresponding hard sphere reference (denoted as HS) run. The only input to these HS systems, containing randomly oriented molecules, were the appropriate density and molecular geometry. In other words, the HS run is a RMC simulation without diffraction data but with all the remaining parameters and constraints (atomic number densities, box lengths, cut-offs and *fnc*-s, see Tables II and III). This kind of comparison proved to be extremely useful in our earlier studies of other (but related) molecular liquids.[8] The present study is based on analyses of partial radial distribution functions and two different kinds of orientational correlation functions (see below, in Section III. C, for details); all these characteristics have been calculated directly from particle coordinates. Averages were taken over 50 independent RMC particle configurations.

**C. Calculation of the orientational correlation functions**

Two different types of distance-dependent correlation functions were calculated for describing the mutual orientations of molecular pairs. First the angle confined by molecular axes, defined by the dipole moments of the molecules, was determined for all molecular pairs, in order to characterize the dipole-dipole correlations, as it has been done for liquids containing molecules with $C_{2v}$[24] and $C_{3v}$[10] symmetry.

The origin of the second calculation is the construction introduced for perfect tetrahedra by Rey[25]. For the present study, we have adapted the construction that was developed for $CXY_3$ liquids[10] (for more examples, see Refs. [26-28]). The $PX_3$ trigonal pyramid (or, in other words, distorted tetrahedron) can be supplemented by a virtual atom, positioned where the lone electron pair of the P atom would be, to form a more regular tetrahedron; the resulting geometrical body is therefore very similar to a $CXY_3$ molecule. Thus two types of ligands are present, as in the $CXY_3$ case. From this point on the original route of Rey is followed: (1) a pair of $PX_3$ molecules is taken; (2) two parallel planes are constructed that contain the centers (phosphorus atoms) of the two molecules and that are perpendicular to the line joining the molecules. Every molecular pair may be classified by a pair of numbers that denote the number of ligands (X atoms or the virtual site) of the two molecular centers that are positioned between the planes.

When atoms connected to the central phosphorous atom (three halides, one virtual atom) are not distinguished then six orientational groups (or 'number-of-ligands', NOL, groups) can be defined[25]. These groups are the corner-to-corner (1:1), corner-to-edge (1:2), edge-to-edge (2:2), corner-to-face (1:3), edge-to-face (2:3) and face-to-face (3:3) orientations. Additionally, the two types of ligands (halides and virtual atom) can be distinguished; in this way 21 NOL subgroups result for a $PX_3$ liquid[10]. The complete list of all subgroups is found in Table IV. This labeling has been performed for all pairs of molecules, resulting in various correlation functions that all depend on the distance between the molecular centers.

Table IV. Division of the original orientation groups of Rey[25] into 'number-of-ligands' subgroups for PX$_3$Y molecular liquids. (X: Cl (PCl$_3$), Br (PBr$_3$), I (PI$_3$). Y: virtual atom.) Subgroups marked by light and dark shading can contribute to parallel and anti-parallel mutual orientations of the dipole vectors, respectively.

| 1:1 | | 1:2(=2:1) | | 2:2 | | 1:3(=3:1) | | 2:3(=3:2) | | 3:3 | |
|---|---|---|---|---|---|---|---|---|---|---|---|
| corner-to-corner | | corner-to-edge | | edge-to-edge | | corner-to-face | | edge-to-face | | face-to-face | |
| X | X | X | X,X | X,X | X,X | X | X,X,X | X,X | X,X,X | X,X,X | X,X,X |
| X | Y | Y | X,X | X,Y | X,X | Y | X,X,X | X,Y | X,X,X | X,X,X | X,X,Y |
| Y | Y | X | X,Y | X,Y | X,Y | X | X,X,Y | X,X | X,X,Y | X,X,Y | X,X,Y |
|   |   | Y | X,Y |   |   | Y | X,X,Y | X,Y | X,X,Y |   |   |

## IV. RESULTS AND DISCUSSION

### A. The total scattering structure factors

The total scattering structure factors (tssf) were calculated from MD, RMC and HS configurations; a comparison with experimental tssf-s is shown in Figures 1 and 2. As the RMC refinement is an 'adjusting/fitting' method, the agreement with the diffraction data is always good: the tssf-s of RMC models all coincide with the experimental curves.

A more interesting point is that the tssf-s computed from the MD particle configurations also show a very good agreement with experimental data. The positions of the maxima and minima are identical; only minor differences are observed in terms of the intensities of the maxima. These small deviations are the features that necessitate a RMC refinement. Nevertheless, it is worth pointing out that this version of the OPLS all-atom force field has been shown to reproduce well the structure of room temperature liquid $PCl_3$ and $PBr_3$, as well as that of molten $PI_3$ just above its melting point, in some way beyond our expectations.

Concerning the comparison of the tssf-s from the HS structural models and the corresponding experimental data sets, differences between the RMC and HS structures are apparent at lower Q values, up to about 8 Å$^{-1}$, see Figures 1 and 2. (Note that beyond that Q value contributions from intramolecular distances dominate, which distances are constrained by the fnc-s in both the HS and RMC calculations; that is, the perfect match of the tssf-s is not surprising.) That is, this simple model is not capable of describing intermolecular correlations quantitatively – apart from liquid $PCl_3$, where the HS system follows the data over the whole Q range nearly perfectly.

Regarding the intermolecular structure, liquid $PCl_3$ thus differs from the other two $PX_3$ liquids since space-filling effects, defined by the molecular shape, dominate the intermolecular interactions. In other words, for this particular liquid, an appropriate knowledge of the molecular shape is sufficient to construct the measurable structural information – although such a statement cannot be made without actually performing the diffraction measurement.

We note that the total scattering structure factor arising from our HS model of $PBr_3$ differs from the one in Ref.3: the sharp spike at 1.4 Å$^{-1}$ is missing. Taking into account that the only difference between these two HS models lies in the values of the minimum interatomic distances allowed, we have performed a series of HS simulations while systematically changing these cut-off-s. We found that the intensity of the sharp maximum in question decreases notably with increasing the P…B minimum intermolecular distance. This finding is in agreement with our earlier statement that results may be rather sensitive to the minimum interatomic distances allowed; this also validates the importance of involving new sources of information (such as MD simulation).

One last observation concerning liquid PBr$_3$: the HS model works noticeably better for the X-ray weighted tssf than for the neutron weighted one. The former is overwhelmed by correlations from prdf-s containing Br: the X-ray data dominantly reflect the packing of the halogen atoms in PX$_3$ liquids. Note, however, that the situation is not always that simple: the best fit to X-ray data is found for PCl$_3$ and the agreement between HS and X-ray diffraction worsens gradually to PI$_3$, which trend implies that the importance of non-steric effects increases with the increasing size of the halogen atom.

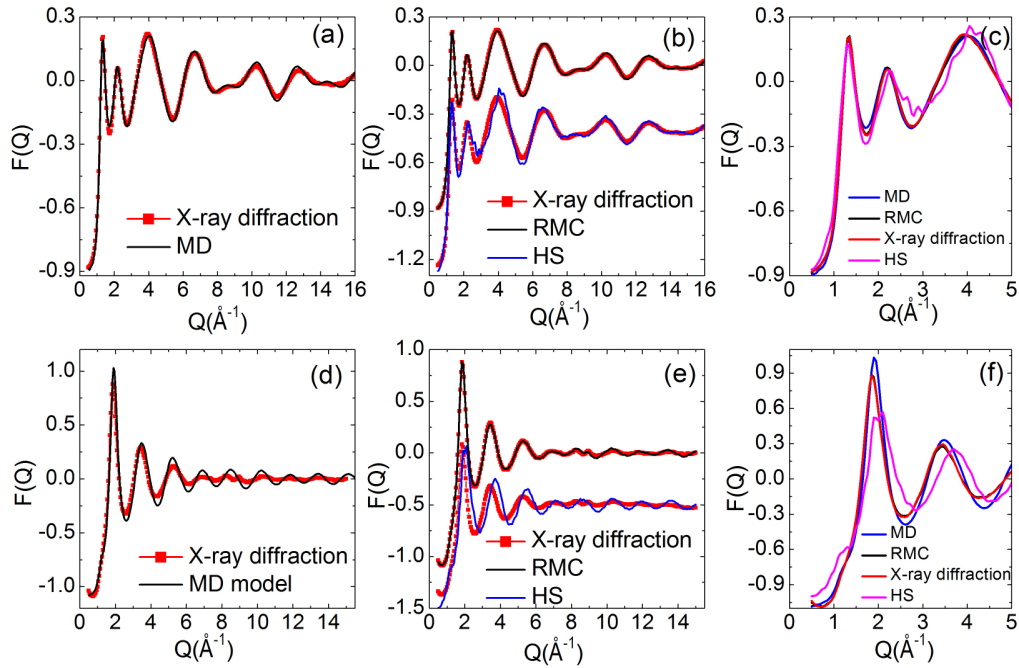

FIG. 1. (a) and (b) X-ray scattering structure factors of liquid PCl$_3$; (d) and (e) X-ray scattering structure factors of molten PI$_3$. Parts (c) and (f) show the three computer models (HS, MD, RMC), together with the measured X-ray diffraction data for PCl$_3$ (c) and PI$_3$ (f), in the range of low Q values.

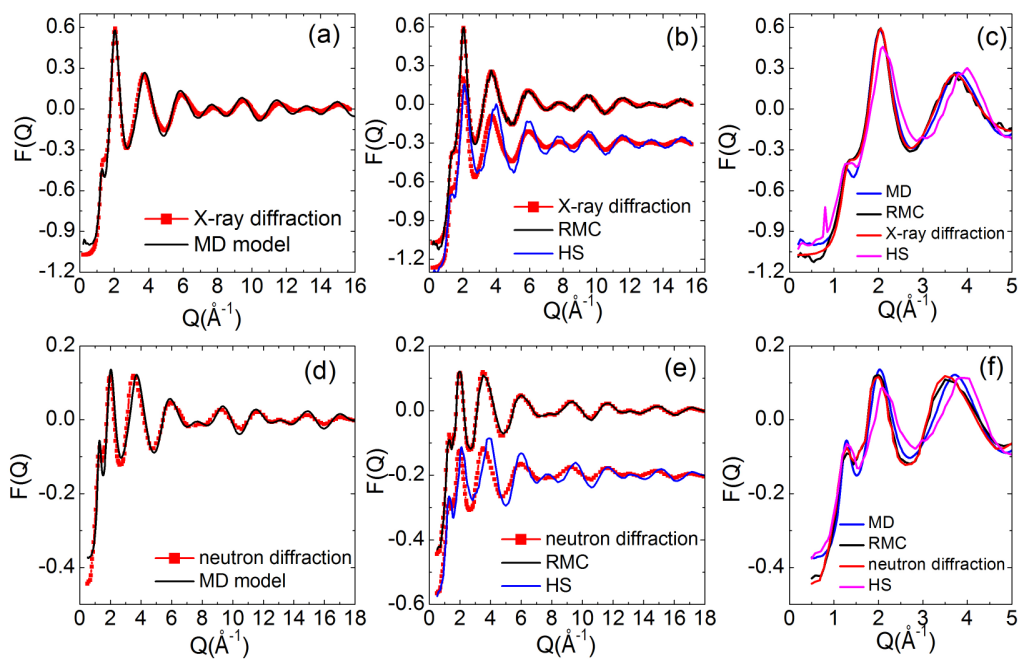

FIG. 2. (a) and (b) X-ray scattering structure factors of liquid PBr$_3$; (c) and (d) neutron scattering structure factors of liquid PBr$_3$. Parts (c) and (f) show the three computer models (HS, MD, RMC), together with the measured X-ray (c) and neutron (f) diffraction data for PBr$_3$, in the range of low Q values.

## B. Partial radial distribution functions

Center-ligand and ligand-ligand partial radial distribution functions from the three simulations (MD, RMC, and HS) are depicted in Figure 3. The first sharp peak (denoted by magenta color) is always the intramolecular contribution: it is virtually identical for the three simulations, confirming the persistence of adequate molecular structures.

Looking at the center-ligand and the ligand-ligand partial radial distribution functions of PCl$_3$ it is obvious that the three structural models (HS, RMC, MD) show close similarities. The first (and only) intermolecular maximum of the P-Cl prdf occurs around 5.3 Å; no oscillations can be detected beyond this distance. Concerning the Cl-Cl prdf, a similar statement can be made: two intermolecular maxima (at 3.8 and 6.4 Å) are located within the distance range of the first P-P peak. This prdf of the HS structural model differs visibly from those of the RMC and MD calculations: the position of the first intermolecular maximum appears at lower $r$ values. The first peak of the HS Cl-Cl prdf is not well separated from the intramolecular part unlike in the cases of the two other structures (RMC and MD). This prdf is also steeper at small distances because the atoms can approach each other down to the (corresponding) distance of closest approach – this is why a 'wall-like' function results. It is important to notice that the Cl-Cl prdf of the MD and RMC structural models clearly show features, e.g. a 'non-wall-like' first intermolecular maximum at larger than the closest allowed Cl-Cl distance, that cannot be related to steric effects only. Interestingly, this is not manifest at the tssf level (cf. Figure 1) – this is one reason why the tssf alone is not sufficient to understand the structure of a liquid.

The Br-Br partial rdf-s look similar to their Cl-Cl counterparts: two intermolecular maxima emerge between 3 and 8 Å and the same observations as above can be made between HS and MD/RMC models.

Concerning the I-I prdfs of liquid $PI_3$, two maxima can also be found, but the intermolecular region is not as well separated from the intramolecular one as for the other two liquids: an iodine atom of one molecule may approach an iodine atom of a neighbouring molecule closer than the intramolecular I-I distance. (We draw here again the attention to the right choice of cut-off distances in the RMC simulations. During this study, setting the I-I and P-Br closest approach distances properly was found to be problematic and in both cases we had to invoke MD results in order to resolve the issue satisfactorily.) As a consequence of the interference between intra- and intermolecular regions and difficulty of choosing the appropriate closest approach distance in RMC, a small extra peak, preceding the intramolecular maximum, may appear in the RMC model that does not occur in the MD model. It seems necessary to tolerate this small artefact in order to reach a good agreement with diffraction data; note that a similar behavior was found for liquid $SnI_4$[28]. Here, this phenomenon can be rationalized by noting that the minimum between the two nearby maxima appears at the intramolecular I-I distance, indicating an imperfect separation of the two (intra- and intermolecular) contributions. Thus, the two apparent maxima correspond to a single, 'real' broad intermolecular peak.

Longer range oscillations can be seen in the P-Br prdf of the RMC model of liquid $PBr_3$; this is particularly striking if one considers that the amplitude of the first intermolecular P-Br maximum is not larger than those of the subsequent peaks. Since these oscillations, and in particular, the one at 7 Å, cannot be found for the HS structure, these features are unambiguously due to diffraction data. We also have to note here that the behavior of this partial is quite different from what has been suggested by previous works on liquid $PBr_3$, especially below 4 Å. The differences may perhaps be ascribed to an inappropriate choice of closest approach values in RMC calculations.

The behavior of the P-P prdf-s (not shown) is typical for the center-center prdf-s of simple molecular liquids that possess a central atom (cf., e.g., Refs. 9, 24, 28). No significant differences can be found between the models. Furthermore, for $PBr_3$ this partial rdf is in good agreement with works published earlier[2,3]. The positions of first maxima and minima of the P-P prdf-s will be important when the orientational correlations will be considered. The maximum positions are 5.5, 5.6 and 5.8 Å, while the first minima appear at 8.0, 8.3 and 9.2 Å for $PCl_3$, $PBr_3$ and $PI_3$, respectively.

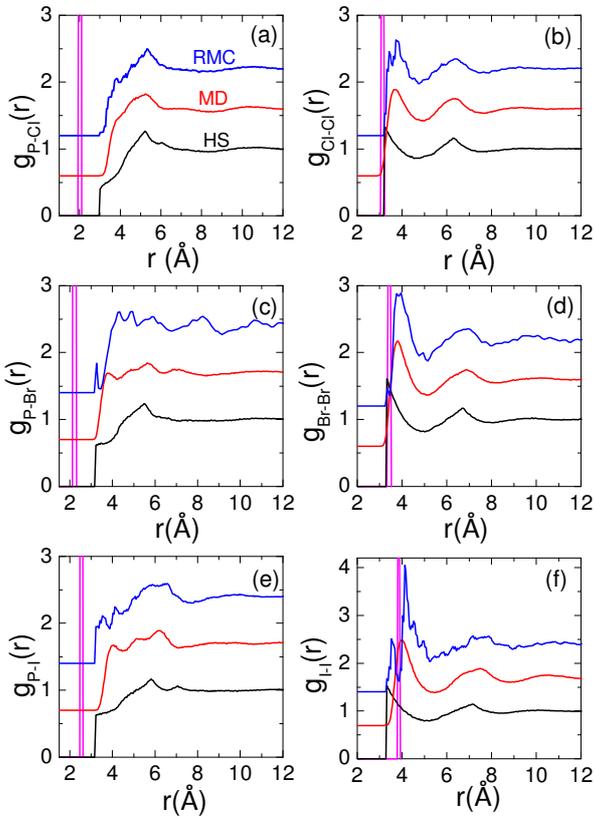

FIG. 3. Partial radial distribution functions for $PX_3$ liquids. (a) P-Cl; (b) Cl-Cl; (c) P-Br; (d) Br-Br; (e) P-I; (f) I-I. Magenta line: intramolecular part for all models; black line: HS; red line: MD; blue line: RMC.

## C. Orientational correlation functions

The two different approaches mentioned in Section 3C are linked in this work (see Figures 4-6). In this way, essential features of the distance dependent dipole-dipole angular correlation function can be explained in terms of specific 'number-of-ligands – number-of-ligands' (NOL—NOL) type contributions. The focus is on the RMC models, although the HS and MD models also are touched upon if relevant.

Selected orientational correlation functions for liquid $PBr_3$, calculated from the RMC particle configurations, are shown in Figure 4. To the best of our knowledge, similar calculations have not yet been performed for this family of liquids[2,3]. The angle between the two dipole vectors has been determined for all molecular pairs and the distribution of their cosines is depicted as a function of the P-P distances in Figure 4a. Concerning the shortest centre-centre distances (within the first P-P coordination shell, up to 8.3 Å) two clear regions of importance emerge (see the red/orange spots in Figure 4a): one around $r$=4.2 Å, $\cos \gamma$ = -1 and another one around $r$=5 Å, $\cos \gamma$=1.

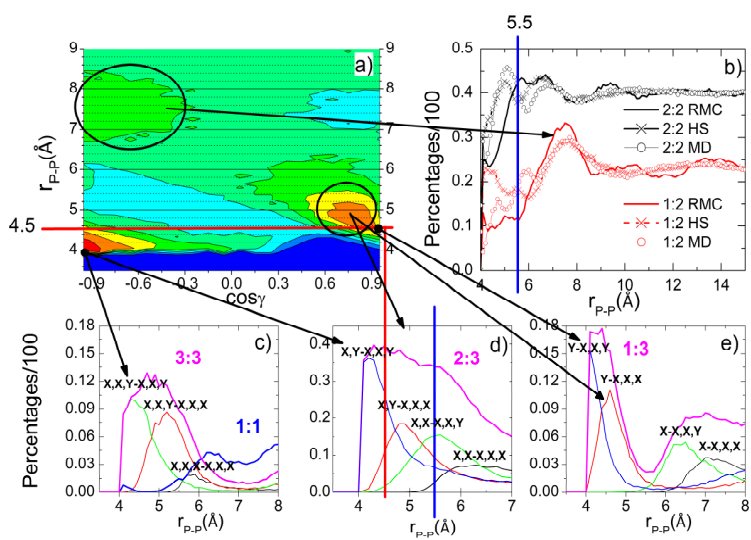

FIG. 4. Orientational correlation functions for PBr$_3$ (see below for terminology). (a) dipole-dipole orientational correlation function (arbitrary relative units; intensity is in increasing order of dark blue, light blue (less than average), light green (average), dark green, yellow, orange, red); (b) 1:2 and 2:2; (c) 1:1 and 3:3 with its subgroups; (d) 2:3 with its subgroups; (e) 1:3 with its subgroups. Subgroup indices: X: Br; Y: virtual atom. The red lines in parts a) and c) mark the upper boundary (at about 4.5 Å) of the distance range within which nearly anti-parallel orientations of the dipole moments are significant (see the red-and-orange spot in part a), lower left corner). The blue lines in parts b) and c) mark the upper boundary (at about 5.5 Å) of the region of nearly parallel orientations (see the orange spot in part a), near the lower right corner). Black arrows connect corresponding regions between the 2D dipole-dipole and the 1D 'NOL' type orientational correlation functions (see text for more details).

As it is mentioned above, in Figure 4 we wish to connect the dipole-dipole and the NOL-NOL type orientational correlation functions in order to provide a more detailed description of the mutual molecular orientations than it was possible previously[8,10]. For this reason

(i) first, the important regions of the 2-dimensional ($r$, cos$\gamma$) distribution were identified (see Figure 4a);

(ii) second, those NOL-NOL correlation functions that possess maxima in the important $r$ regions have been found (see Figures 4c, 4d and 4e; the black arrows connect corresponding regions of the dipole-dipole and the NOL-NOL correlation functions);

(iii) finally, the relevant NOL-NOL subgroups had to be selected that may be at the origin of the dipole-dipole mutual orientation identified under (i) (see Table IV for the assignation).

The highest intensities in Figure 4a, at around 4.2 Å, belong to the antiparallel orientation (cos $\gamma$ = - 1). Antiparallel dipolar arrangements are naturally connected with the 3:3 and 1:1 groups (Fig. 4c): the subgroups X,X,X-X,X,X of the former and Y-Y of the latter are trivial examples. Most of the X,X,Y-X,X,Y (3:3 type) pairs possibly also form (almost) antiparallel mutual molecular arrangements (remember: X: bromine; Y: virtual atom). A particular subgroup of the 2:3 group (Fig. 4d), the X,Y-X,X,Y orientation, may also be seen as an antiparallel arrangement. As evidenced by Fig. 4c, the 1:1 group hardly contributes over this distance range: it reaches a ratio of about 5% beyond 6 Å. Considering the 3:3 orientation (Fig. 4c), the contribution of the X,X,Y-X,X,Y arrangements, that may be anti-parallel, is significant below 4.8 Å. (The other main contributor from the 3:3 group, up to 6 Å, is the X,X,X-X,X,Y arrangement that cannot be anti-parallel). Beyond the P-P distance of about 6 Å the share of the 3:3 group (Figure 4c) decreases steeply and ends up below the level of the 1:1 group.

Comparing the three structural models (MD, RMC, HS) in the region below 4.5 Å, the contribution of the X,X,X-X,X,Y subgroup in both the HS (4%) and the MD (5.2%) models is much lower than in the RMC model (8.7%). The share of the X,X,Y-X,X,Y subgroup is below 4 % in the HS system; in contrast, in the MD and RMC models X,X,Y-X,X,Y pairs are much more abundant, with an occurrence of about 10 %. That is, none of the important orientations of the 3:3 group can be explained by the HS model (i.e., by steric effects).

The parallel orientation, with $\cos \gamma = 1$ (see Figure 4a), may correspond to the 1:3 group (Figure 4e; see also Table IV): the clear case is Y-X,X,X , which arrangement necessitates that the two dipolar vectors are on the same line and point to the same direction (cf. the 'Apollo'-model[8,10]). The X-X,X,Y subgroup (Figure 4e) may also contain a large number of pairs that arrange (almost) parallel with respect to each other. There are slightly more molecular pairs in the 1:3 group (at its peak the curve reaches 17 %, see Figure 4e) over a narrower region than in the 3:3 group (cf. Figure 4c). The maximum for the Y-X,X,X subgroup appears at around 4.8 Å, which value matches exactly the position of the 'parallel' peak of the dipole-dipole orientational correlation function (see Fig. 4a). There is at least one more recognizable contributor to the 'parallel' peak in its high angle part ($\cos \gamma = 0.7$): the correlation function for the X,Y-X,X,X subgroup of the 2:3 group (Fig. 4d) peaks at 5 Å.

The two important regions mentioned above are well separated (see Figure 4a for liquid PBr$_3$; the horizontal red line symbolizes the line of separation between the regions). In summary, molecular pairs whose centre-centre distance is less than 4.5 Å are closer to the antiparallel arrangement, while above 4.5 Å (and up to about 5.5 Å) the parallel-like orientation is more likely.

Concerning the region of centre-centre distances beyond 5.5 Å, 2:2 orientations (Fig. 4b) are dominant: this is evidenced by the good match between the positions of the main peak of the corresponding NOL-NOL correlation function (Figure 4b, curves in red) and the broad maximum on the 'anti-parallel side' of Figure 4a. Additionally, we draw the attention to the contribution of the 1:2 group (Fig. 4b, black lines), which results in somewhat higher than average intensities in the dipole-dipole orientational correlation map around 7.5 Å (Fig. 4a). This peak is also found in the HS and MD systems.

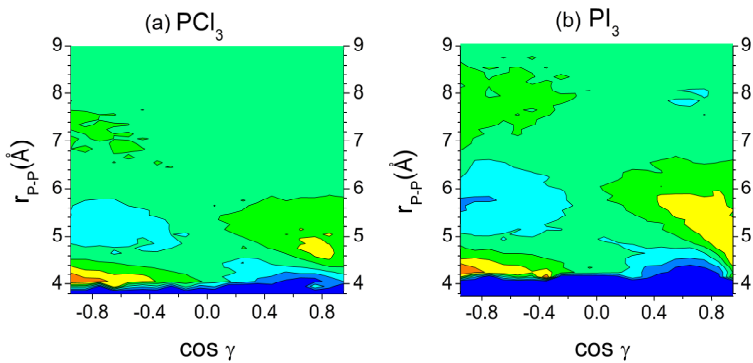

FIG. 5. Dipole-dipole orientational correlation functions. (a) $PCl_3$_RMC model; (b) $PI_3$_RMC model. (Arbitrary relative units; intensity is in increasing order of dark blue, light blue (less than average), light green (average), dark green, yellow, orange.)

Turning now to the other two liquids, the dipole-dipole orientational correlation functions for $PCl_3$ and $PI_3$ can be found in Figures 5a and 5b. At first sight these plots resemble to that of $PBr_3$, although with distinct alterations. Firstly, the two maxima are found at roughly the same distance and cosine values, but not with the same intensities as $PBr_3$. The displacements of the spots of high probability can largely be ascribed to the different molecular sizes (although differences between intermolecular interactions may well be responsible, as well). However, they correspond approximately to the same orientations as found for $PBr_3$ (see below and Fig. 6).

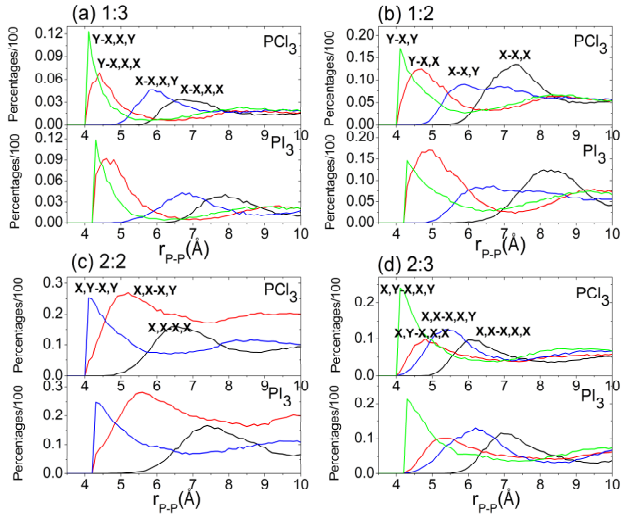

FIG 6. Orientational correlation functions (a) Subgroups of 1:3; (b) subgroups of 1:2; (c) subgroups of 2:2; (d) subgroups of 2:3. Subgroup indices: X: Cl ($PCl_3$), I ($PI_3$); Y: virtual atom.

Molecular pairs with more or less anti-parallel mutual orientations belong to groups 2:3 (subgroup X,Y-X,X,Y), 2:2 (subgroup X,Y-X,Y) and 1:2 (subgroup Y-X,Y), as evidenced by the match between peak positions of the dipole-dipole and the NOL-NOL correlation functions. Discrepancies between $PCl_3$ and $PI_3$ manifest only in terms of fine details, like that the role of the X,Y-X,X,Y and Y-X,Y subgroups is more explicit in liquid phosphorus trichloride (see Fig. 6). Parallel arrangements are most likely to be caused mainly by groups 1:3 (subgroup Y-X,X,X) and 1:2 (subgroup Y-X,X); subgroup X,Y-X,X,X (of group 2:3) may well appear in liquid $PI_3$ but not in $PCl_3$ (see Fig. 6).

The main difference that cannot be explained by the different molecular shape is the occurrence of (near-) parallel orientations: in liquid $PI_3$ (Fig. 5 (b)), they are significantly less favoured than in liquid $PBr_3$, whereas they are almost missing from $PCl_3$. This tendency is the strongest in the MD systems. Similarly, the magnitude of the antiparallel orientations is the smallest in $PCl_3$. Concerning variations among the three computational models (MD, RMC and HS), the marked differences between them present in liquid $PBr_3$ are hardly found for the other two liquids ($PCl_3$ and $PI_3$). This is not unexpected in the case of $PCl_3$, where similar statements were concluded for the three systems during the analyses of tssf-s and prdf-s. On the other hand, for liquid $PI_3$, deviations between HS and RMC models in terms of the tssf-s suggested larger differences between the corresponding orientational correlation functions.

## V. SUMMARY AND CONCLUSIONS

New synchrotron X-ray diffraction experiments have been conducted for the $PX_3$ (X: Cl, Br, I) family of molecular liquids and their total scattering structure factors have been determined.

Molecular dynamics simulations in the canonical ensemble, applying the OPLS all-atom force-field, have then been performed for the three liquids. Total scattering structure factors calculated from the MD trajectories match the measured X-ray data extremely well, almost within the (estimated) experimental uncertainties.

Data from the new X-ray diffraction measurements have been used to drive subsequent Reverse Monte Carlo calculations that were started from the final particle configurations of the MD simulations. The available neutron diffraction data for liquid $PBr_3$ have been also utilized. In parallel, hard sphere (HS) Monte Carlo (technically, RMC without any experimental data) calculations have also been conducted in order to separate packing effects from genuine intermolecular interactions like the dipole—dipole one. Thus our conclusions are based on comparisons between the following nine (3x3) structural models: $PCl_3$_MD, $PCl_3$_HS, $PCl_3$_RMC, $PBr_3$_MD, $PBr_3$_HS, $PBr_3$_RMC, $PI_3$_MD, $PI_3$_HS and $PI_3$_RMC.

For a detailed characterization of the structure, total scattering structure factors, partial radial distribution functions, and two different orientational correlation functions have been calculated. The following findings emerged from the present study:

(1) The most likely orientation of two neighbouring $PX_3$ molecules at the closest P-P (i.e., molecular center—center) distances is the antiparallel-like arrangement, which is mostly composed of the X,Y-X,X,Y, X,Y-X,Y, X,X,Y-X,X,Y, X,X,X-X,X,X and Y-X,Y subgroups (see Table IV).

(2) Parallel-like orientations become typical at distances between 4.5 and about 6 Å and originate mainly from the Y-X,X,X, Y-X,X and X,Y-X,X,X subgroups.

(3) At larger centre-centre distances easily recognizable dipole arrangements disappear due to increasing intensity from subgroups of the 2:2 group.

(4) At all levels of the structural analyses, from total scattering structure factors to orientational correlation functions, the three models (MD, HS and RMC) of $PCl_3$ show very similar, nearly indistinguishable features. The structure of liquid $PI_3$ appears to be more similar to the structure of $PCl_3$ than to that of $PBr_3$, which appears to be the most structured of the three liquids. Steric effects are therefore insufficient for explaining the main characteristics of $PBr_3$ and it is conjectured that dipole—dipole interactions play the most important role within the $PX_3$ family.

Finally, we wish to stress that for clarifying the microscopic structure of these molecular liquids on the basis of diffraction experiments, Molecular Dynamics simulations are indeed an indispensable aid.

**Acknowledgment**

Financial support was provided by the Hungarian National Basic Research Fund (OTKA), via Grant No. 83529. X-ray diffraction experiments in SPring-8 (Hyogo, Japan) were carried out under proposal No. 2010B1085. LT is grateful to JSPS for a postdoctoral fellowship and to his host Dr. S. Kohara, who made it possible to carry out the research at the Japan Synchrotron Radiation Research Institute between 2009 and 2011. LT and LP are grateful to H. Iwamoto for his help during the sample preparation in the Chemistry Preparation Room of SPring-8. LT is grateful to Á. Temleitner for useful advices concerning the proper sealing procedure of the sample capillaries.

**References**

[1] R. Enjalbert, J.-M. Savariault, J.-P. Legros, C.R. Acad. Sc. Paris, Ser. C **290**, 239 (1980); R. Enjalbert, J. Galy, Acta Cryst. **B35**, 546 (1979); E. T. Lance, J. M. Haschke, D. R. Peacor, Inorg. Chem. **15**, 780 (1976)

[2] M. Misawa, T. Fukunaga and K. Suzuki, J. Chem. Phys. **92**, 5486 (1990)

[3] B. J. Gabrys, L. Pusztai and D. G. Pettifor, J. Phys. Condensed Matter **19**, 335205 (2007)

[4] M. P. Allen and D. Tildesley, Computer Simulation of Liquids (Clarendon Press; Oxford 1987).

[5] R. L. McGreevy and L. Pusztai, Mol. Simul. **1**, 359. (1988).

[6] O. Gereben and L. Pusztai, J. Phys. Chem. B **116**, 9114 (2012).

[7] D. A. Keen, J. Appl. Cryst. **34**, 172 (2001).

[8] S. Pothoczki, L. Temleitner and L. Pusztai (2012) Determination of Molecular Orientational Correlations in Disordered Systems from Diffraction Data, in Advances in Chemical Physics, Volume 150 (eds S. A. Rice and A. R. Dinner), John Wiley & Sons, Inc., Hoboken, NJ, USA.

[9] S. Pothoczki and L. Pusztai, J. Mol. Liq. **145**, 38 (2009).

[10] S. Pothoczki, L. Temleitner and L. Pusztai, J. Chem. Phys. **134**, 044521 (2011).

[11] M. Isshiki, Y. Ohishi, S. Goto, K. Takeshita, T., Nuc. Instr. and Methods. in Phys. Res. **A467-8**, 663-666 (2001).

[12] F. E. E. Germann, R. N. Traxler, J. Am. Chem. Soc. **49**, 307 (1927).

[13] S. Kohara, M. Itou, K. Suzuya, Y. Inamura, Y. Sakurai, Y.Ohishi and Masaki Takata, J. Phys.: Condens. Matter **19**, 506101 (2007).

[14] S. Sasaki, KEK Report **88**-14 (1989).

[15] O. Gereben and L. Pusztai, Phys. Rev. B 51, 5768 (1995).

[16] D. van der Spoel, E.Lindahl, B. Hess, G. Groenhof, A. E. Mark and H. J. C. Berendsen, J. Comput. Chem. **26**, 1701 (2005); B. Hess, C. Kutzner, D. van der Spoel and E. Lindahl, J. Chem. Theory Comput. **4**, 435 (2008) ( http://www.gromacs.org).

[17] W. L. Jorgensen, D. S. Maxwell and J. Tirado-Rives, J. Am. Chem. Soc. **118**, 11225 (1996).

[18] B. Hess, H. Bekker, H. J. C. Berendsen and J. G. E. M. Fraaije, J. Comput. Chem. **18**, 1463 (1997).

[19] H. J. C. Berendsen, J. P. M. Postma, A. DiNola and J. R. Haak, J. Chem. Phys. **81**, 3684 (1984).

[20] O. Gereben, P. Jóvári, L. Temleitner, and L. Pusztai, J. Optoel. Adv. Mater. **9**, 3021 (2007).

[21] R. L. McGreevy, J. Phys.: Condens. Matter **13**, R877 (2001).

[22] G. Evrard and L. Pusztai, J. Phys: Condens. Matter **17,** S1 (2005).

[23] O. Gereben and L. Pusztai; Journal of Computational Chemistry **33**, 2285 (2012)

[24] S. Pothoczki, L. Temleitner and L. Pusztai, J. Chem. Phys. **132**, 164511 (2010).

[25] R. Rey, J. Chem. Phys. **126,** 164506 (2007)

[26] N. B. Caballero, M. Zuriaga, M. Carignano and Pablo Serra, J. Chem. Phys. **136**, 094515 (2012).

[27] N. B. Caballero, M. Zuriaga, M. A. Carignano and Pablo Serra, Chem. Phys. Lett. **585**, 69-73 (2013).

[28] L. Pusztai, Sz. Pothoczki and S. Kohara, J. Chem. Phys. 129, 064509 (2008).